\documentstyle[preprint,eqsecnum,aps]{revtex}
\begin{document}
\draft
\title{On matching conditions for cosmological perturbations}
\author{Nathalie Deruelle}
\address{D\'epartement d'Astrophysique Relativiste et de Cosmologie\\
Centre National de la Recherche Scientifique,\\ Observatoire de Paris,
92195 Meudon Cedex, France,  and\\
Department of Applied Mathematics and Theoretical Physics\\
University of Cambridge, Silver Street Cambridge CB3 9EW, England}
\author{V.F. Mukhanov\cite{Muk}}
\address{Institut f\"ur Theoretishe Physik,\\ ETH, H\"onggerberg,
 CH-8001 Z\"urich, Switzerland, and, \\
D\'epartement d'Astrophysique Relativiste et de Cosmologie\\
Centre National de la Recherche Scientifique,\\ Observatoire de Paris,
92195 Meudon Cedex, France}
\date{\today}
\maketitle
\begin{abstract}
  We derive the matching conditions for cosmological perturbations in
a Friedmann Universe where the equation of state undergoes a sharp jump,
for instance as a result of a phase transition. The physics of
the transition which is needed to follow the fate of the perturbations is
clarified. We dissipate misleading statements made recently in the
literature \cite{2} concerning the predictions of the primordial
fluctuations from inflation and confirm standard results. Applications
to string cosmology are considered.
\end{abstract}

\mediumtext

\section{Introduction}
\label{sec:1}

The evolution of metric perturbations in a Friedmann Universe is well
known (see, e.g. \cite{1} and refs therein). They can be classified into
scalar perturbations, which couple to the matter density inhomogeneities,
 and freely propagating gravitational waves (we shall ignore here the
vectorial perturbations). In inflationary models [3] the perturbations of
both types are produced from primordial quantum fluctuations, and their
spectra after inflation are nearly scale invariant [4,5]. However the
amplitudes of the gravitational waves and scalar perturbations are
significantly different at all scales after the transition from
the inflationary to the matter dominated stage.

Let us, for instance
consider a Universe which evolves from a phase (--) with the scale factor
$a_- \propto t^{p_-}$ to a phase (+) such that $a_+ \propto t^{p_+}$.
 If the transition from the (--) and to the (+) phase happens at time
$t_{t}$, then the ratio of the amplitude $h^+$ of the long wavelength
gravity waves some time after the transition  $(t\gg t_{t})$ and
directly before it $h^-$ (at $t=t_{t} -0)$ is:
 \begin{equation}
   {h^+\over h_-} \to O(1) \label{eq:1.1}
 \end{equation}
while for the gravitational potential $\Phi$, which characterizes scalar
metric perturbations, the result is
 \begin{equation}
   {\Phi^+\over \Phi^-} \to {1+p_-\over 1+p_+}\ . \label{eq:1.2}
 \end{equation}
We see that for a strongly inflating (--) phase $(p_- \gg 1)$ the
amplification of density perturbations can be very big whereas gravity
waves are not amplified.

 This standard result has however been challenged recently in \cite{2}.

 We think that there is nothing strange about this difference in the
amplification of the gravity waves and density perturbations.
Of course, eq.~(\ref{eq:1.2}) doesn't imply that
the gravitational potential jumps at the moment of transition. What
happens is that the gravitational potential $\Phi$ which can be small
immediately before and after the transition is distributed at $t
=t_{t} +0$ into two modes, both very large in amplitude, one which
decays, the other yielding (\ref{eq:1.2}) (the derivative $\dot\Phi$ of the
gravitational potential on the other hand can strongly jump at $t_{t}$
in distinction from the gravity waves for which $\dot h$ is continuous).
As we will show the result depends on the physics of the transition.

 In order to make this statement quantitative we derive the matching
conditions for cosmological perturbations in models with sharp transitions.
The results will be applied to the simplified model of inflation
considered in \cite{2} which uses a sharp jump of the equation of state
at the end of inflation instead of evolving it rapidly but smoothly as
in e.g. \cite{1} (see also e.g. \cite{6,7}). We think that the results
obtained in \cite{2} imply the strongly scale dependent spectrum for
scalar perturbation in the relevant for COBE scales.  We do not agree
with the author of paper \cite{2} who concluded from his consideration
that the spectrum is flat taking the parameter $\gamma_1$ to be equal 2
on p.~7166, while it should be taken $\gamma_1 \ll 1$.  It will be shown
in Sect.~\ref{sec:5.2} that the root for the misleading result \cite{2}
lies in inappropriate matching conditions which have no physical
justification.  We on the other hand shall make the matching on
hypersurfaces of constant energy for physical reasons and clarify and
{\it confirm the standard results} (\ref{eq:1.1})--(\ref{eq:1.2}).

The considerations in this paper actually go beyond clarifying this point.
They can be applied to study the evolution of
cosmological perturbations in models where the transition from one
stage of evolution to an other is not completely specified. We clarify
the physical features of the phase transition which are of importance
for the evolution of cosmological perturbations. As a particular example
we consider the fate of the growing mode in string cosmology after
transition from superinflation to the Friedmann era (see
Sect.~\ref{sec:5}).

 The paper is organized as follows. In Section~\ref{sec:2} we recall the
standard results. In Section~\ref{sec:3} the general matching conditions
in models with sharp transitions are derived.  We also discuss there the
choice of the hypersurface of transition. In Section~\ref{sec:4} we write
the matching conditions in two particular coordinate systems: the
longitudinal and the synchronous reference frame used in \cite{2}.
Finally, in Section~\ref{sec:5}, we apply our results to concrete
examples.  Conclusions are drawn in Section~\ref{sec:6}.

\section{Background and perturbations} \label{sec:2}

Consider a quasi-isotropic spatially flat Friedmann Universe with metric
 \begin{equation}
  ds^2 = a^2 \{ (1+2\phi) d\eta^2 - 2B_{,i} dx^i d\eta - [(1-2\psi)
   \delta_{ik} + 2E_{,ik} + h_{ik}] dx^i dx^k \} \label{eq:2.1}
 \end{equation}
where $a(\eta)$ is the scale factor, a comma means the derivative
with respect to spatial coordinates and the four functions $\phi$,
$\psi$, $B$ and $E$ characterize the metric perturbations of scalar
type. The tensor $h_{ik}$, which is taken transverse and traceless
$(h^{ik}_{\ ,k} = h^i_i = 0)$ corresponds to gravity waves (all spatial
indices will be raised with $\delta^{ik}$). We ignore here the vector
perturbations. Under coordinate transformations $\eta \to \tilde\eta
=\eta +\xi^0$, $x^i \to \tilde x^i = x^i +\xi^{,i}$, the scalar metric
perturbations transform as:
 \begin{eqnarray}
  &\phi \to \tilde \phi = \phi - {\cal H} \xi^0 - \xi^{0'} \quad ; \quad
 &B \to \tilde B = B + \xi^0 - \xi^{0'} \ ; \nonumber \\
  &\psi \to \tilde \psi = \psi - {\cal H} \xi^0 \quad ; \quad
 &E \to \tilde E = E - \xi\ ; \label{eq:2.2}
\end{eqnarray}
where ${\cal H} =a'/a =Ha$ and a prime denotes a derivative with respect
to conformal time $\eta$.

{}From (\ref{eq:2.2}) we see that two simple gauge invariant quantities,
characterizing the scalar perturbations, can be built, e.g. \cite{8,2}
 \begin{equation}
  \Phi = \phi + {1\over a} [(B-E') a]' \ , \ \Psi = \psi - {\cal H}
   (B-E')\ . \label{eq:2.3}
\end{equation}
In the longitudinal gauge $(B_\ell = E_\ell =0)$ these gauge invariant
variables coincide with the metric perturbations.  Therefore the
calculations in this coordinate system are completely the same as
those in terms of the gauge invariant variables $\Phi$ and $\Psi$.  As
for the gravity waves they are gauge invariant:  $\tilde h_{ij} =
h_{ij}$.

 If we consider the perturbations with scales much bigger than the Hubble
radius $H^{-1}$, then the solution of Einstein equations in the long
wavelength approximation can be written down as (see e.g. \cite{1}):
\begin{equation}
 h =  D_g + S_g \int {dt\over a^3} \label{eq:2.4}
\end{equation}
for gravity waves and
\begin{equation}
  \Phi = \Psi = S_s {H\over a} +  D_s \left( {1\over a} \int
   adt \right)^. \label{eq:2.5}
\end{equation}
for scalar perturbations in the case of matter an adiabatic perfect
fluid or a scalar field.
Here a dot denotes a derivative with respect to cosmic time $t=\int
ad\eta$, and $H=\dot a/a$. The constants of integration $D$,
$S$ are fixed by physical considerations, e.g. by imposing that the
perturbations be in their quantum ground state when they are inside the
Hubble radius during an inflationary era.
We note that the solution (\ref{eq:2.5}) can be obtained if we integrate
the equation for the quantity (see, for instance \cite{1,5,9,10})
 \begin{equation}
   \zeta = \Phi + {{\cal H}\over {\cal H}^2 - {\cal H}'}
  ( \Phi' + {\cal H} \Phi ) \label{eq:2.6}
\end{equation}
which is ``conserved'' in the long wavelength approximation.

 Let us pause onto the model described in the Introduction, where the scale
factor evolves as $a\propto t^{p_-}$ before the transition and as
$a\propto t^{p_+}$ after it. Then using eqs.~(\ref{eq:2.4}), (\ref{eq:2.5})
one gets that $h^- \to D_g$, $\Phi^- \to D_s / (1+p_-)$ and after the
transition after the subdominant mode has decayed:  $h^+\to D_g$,
$\Phi^+ \to D_s / (1+p_+)$.  From here immediately follow the results
(\ref{eq:1.1}), (\ref{eq:1.2}).  Since during inflation $1+p_- \propto
H^2/\dot H \gg 1$ we see that the scalar perturbations can be very
strongly amplified in the course of the transition.

The standard derivation described above was criticized in \cite{11}
on the grounds that the long wavelength approximation used breaks down
for scalar perturbations of cosmologically relevant scales at the moment
of the transition. It is true that the condition of applicability of
the long wavelength approximation for scalar perturbations can be
significantly different from simply imposing that the wavelength
be bigger than the Hubble scale (see \cite{1} and \cite{7}). Therefore
we believe that it is quite useful to clarify how the perturbations
evolve in models with a sharp change of the equation of state and
discuss the features of the physics of the transition which are relevant
for cosmological perturbations.

\section{Matching conditions}  \label{sec:3}

 Suppose that the stress-energy tensor, which governs the evolution of
the metric (\ref{eq:2.1}) via Einstein's equations undergoes a finite
discontinuity on a spacelike hypersurface $\Sigma$ determined by the
equation $q(\eta , x^i) ={\rm const.}$, where $q$ is 4-scalar.  In a
slightly inhomogeneous Universe it can be decomposed into a
``homogeneous'' part $q_0(\eta)$ plus a small inhomogeneity $\delta
q(\eta , x^i)$, that is $q=q_0+ \delta q$.  Under coordinate
transformations $\eta =\tilde \eta = \eta +\xi^0$ etc..., $\delta q$
changes as
\begin{equation}
 \delta q \to \widetilde\delta q = \delta q - q'_0 \xi^0\ . \label{eq:3.1}
\end{equation}
We shall consider below specific examples for the function $q$. The
hypersurface $\Sigma$ divides the manifold into two regions, $(-)$ and
(+) which should be matched in such a way that the induced 3-metric
on $\Sigma$ and its extrinsic curvature be continuous (in the absence of
surface layers) \cite{12}.  The simplest way to do the calculation is to
go to the ``tilde coordinate system'':  $\eta \to \tilde\eta$, $x^i\to
\tilde x^i$, where the equation for $\Sigma$ becomes $\tilde \eta = {\rm
const.}$.  It means that in this coordinate system $\widetilde{\delta q}
= 0$ and eq.~(\ref{eq:3.1}) determines the transformation:
\begin{equation}
  \xi^0 = {\delta q\over q_0'} \quad , \quad \xi -{\rm arbitrary} \ .
  \label{eq:3.2}
\end{equation}
{}From this, one first immediately gets that for the background and
gravity waves the matching conditions impose that the scale factor
$a$, $h_{ik}$ and their first time derivatives be continuous on
$\Sigma$ in any coordinate system. For the scalar perturbations we can
easily calculate the extrinsic curvature $\widetilde{\delta K}_j^i$ in
the ``tilde coordinate system'' and the matching conditions for them
read:
\begin{mathletters}
\label{eq:3.3}
\begin{eqnarray}
&&[\tilde \psi]_{\pm} = [\tilde E]_{\pm} = 0\ , \label{eq:3.3a}\\
&&[\widetilde{\delta K}^i_j]_{\pm} = -{1\over a} [\delta^i_j ({\cal H}
 \tilde\phi + \tilde\psi ') + (\tilde B-\tilde E')^{,i}_{\ ,j}]_{\pm}
  = 0 \ ,   \label{eq:3.3b}
\end{eqnarray}
\end{mathletters}
where $[\tilde \psi]_{\pm} \equiv \tilde\psi_+ -\tilde\psi_-$ etc.
{}From eq. (\ref{eq:3.3b}) it follows that $({\cal H}\tilde \phi + \tilde
\psi')$ and $(\tilde B -\tilde E')$ should be separatly continuous.
Expressing $\tilde\psi$, $\tilde B$ etc... in terms of $\psi$, $B$ etc...
in the original coordinate system with the help of (\ref{eq:2.2}) where
$\xi^0$ should be taken from (\ref{eq:3.2}), we finally get the matching
conditions on $\Sigma$: $q_0+\delta q= {\rm const.}$ in an arbitrary
coordinate system:
\begin{mathletters}
\label{eq:3.4}
\begin{eqnarray}
 &&[\psi + {\cal H} \delta q / q_0']_{\pm} = 0\ , \label{eq:3.4a}\\
 &&[ E -\xi]_{\pm} = 0 \ , \label{eq:3.4b}\\
 && [B-E' + \delta q /q'_0]_{\pm} = 0\ , \label{eq:3.4c} \\
 && [{\cal H}\phi +\psi' + ({\cal H}' -{\cal H}^2)
  \delta q/q_0']_{\pm} = 0 \ .   \label{eq:3.4d}
\end{eqnarray}
\end{mathletters}
Eq. (\ref{eq:3.4b}) is empty since $\xi$ can always be chosen so that
it is satisfied.

Now it is time to specify $\Sigma$, that is the scalar $q$. Suppose matter is
an adiabatic perfect fluid with stress-energy tensor $T^\mu_\nu =
(p+\varepsilon) u^\mu u_\nu - p\delta^\mu_\nu$, and that the equation
of state $p =p(\varepsilon)$ undergoes a sudden change at some moment
of time. Then it is clear that on $\Sigma$ the energy density
$\varepsilon$ should be constant (that is, $\Sigma$: $q\equiv
\varepsilon_0 +\delta \varepsilon ={\rm const.})$, since the pressure
depends only on $\varepsilon$. The background energy density
$\varepsilon_0$ and its perturbations $\delta\varepsilon$ can be
expressed in terms of the scale factor $a$, and the metric perturbations
$\phi$, $\psi$ etc... via the linearized $0-i$ Einstein eqs. and thus
the eqs.~(\ref{eq:3.4}) can be written down entirely in terms of the metric
perturbations. We will do it in next Section for various gauges.

If matter is a scalar field $\varphi$ with stress-energy tensor $T^\mu_\nu
=\varphi^{,\mu} \varphi_{,\nu} - \delta^\mu_\nu ({1\over 2} \varphi^{,\lambda}
\varphi_{,\lambda} - V(\varphi))$ which suddenly decays into ultrarelativistic
particles then $\Sigma$ should be taken to be
the hypersurface $\varepsilon = {1\over 2} \varphi^{,\lambda}
\varphi_{,\lambda} + V(\varphi) = {\rm const.}$, where $\varepsilon$
is the ``effective'' energy of a scalar field considered as a ``perfect
fluid'' with 4-velocitie $u_\mu = \varphi_{,\mu} /
(\varphi^{,\lambda} \varphi_{,\lambda})^{1/2}$.

\section{Longitudinal VS. synchronous gauge} \label{sec:4}

\subsection{Longitudinal gauge}  \label{sec:4.1}

The longitudinal gauge conditions: $E_\ell = B_\ell = 0$, define a unique
coordinate system. In this coordinate system the matching conditions
(\ref{eq:2.4}) on the hypersurface $\Sigma :  \varepsilon_0 + \delta
\varepsilon_\ell =$ const simplify to:
\begin{eqnarray}
&& [{\delta \varepsilon_\ell\over \varepsilon_0'}]_\pm = 0 , \quad
[\psi_\ell]_\pm = 0 , \\
&& [{\cal H} \phi_\ell + \psi'_\ell + {\cal H}' \delta \varepsilon_\ell/
\varepsilon_0]_\pm = 0 \ . \label{eq:4.1}
\end{eqnarray}
Now we will use the background $0 - 0$ Einstein equation to express
$\varepsilon_0$ and $\varepsilon_0'$ in terms of the scale factor and
its derivatives:
\begin{equation}
  \varepsilon_0 = {3{\cal H}^2\over \kappa a^2}\ ,\qquad
  \varepsilon_0' = {6\over \kappa}{{\cal
      H}\over a^2} ({\cal H}' - {\cal H}^2)\ , \label{eq:4.2}
\end{equation}
where $\kappa = 8\pi G$, and the $i-k$ $(i\not= k)$ (if is supposed that
$\delta T^i_k =0$ for $i\not= k$ together with $0-0$ linearized
Einstein equations in the longitudinal gauge (see Sec. in \cite{1})
\begin{equation}
 \psi_\ell = \phi_\ell ,\quad\delta\varepsilon_\ell = {6\over \kappa a^2}
 \left[{1\over 3} \Delta \phi_\ell - {\cal H} ({\cal H} \phi_\ell +
 \phi_\ell')\right] \ . \label{eq:4.3}
\end{equation}

Then substituting these eqs. in (\ref{eq:4.1}) we arrive at the following
two independent matching conditions (the third one becomes redundant):
\begin{mathletters}
\label{eq:4.4}
\begin{eqnarray}
&& [\phi_\ell]_\pm = 0 , \label{eq:4.4a} \\
&& \left[\zeta_\ell + {1\over 3} {\Delta \phi_\ell\over {\cal H}'
  - {\cal H}^2} \right]_\pm =0 \ , \label{eq:4.4b}
\end{eqnarray}
\end{mathletters}
where $\zeta_\ell$ is the ``conserved'' quantity defined by (\ref{eq:2.6})
(recall that $\Phi = \phi_\ell$). The second term proportional to $\Delta
\phi_\ell$ in (\ref{eq:4.4b}) can be neglected for the perturbations
with scales bigger than the Hubble scale $H^{-1}$.  Then we see that for
these long wavelength perturbations the matching condition (\ref{eq:4.4b})
reduces to the widely used conservation law for $\zeta_\ell$.  It is
worth to mention that for long wavelength perturbations the continuity
of $\phi_\ell$ and $\zeta_\ell$ also implies that the 3-velocity
$u_\ell^i$ should be continous on $\Sigma$.

We conclude this paragraph by noting that all Eqs.
(\ref{eq:3.1})-(\ref{eq:3.4}) can be
turned into relations between gauge invariant quantities by changing
$\psi_\ell$, into $\Psi$, $\phi_\ell$ into $\Phi$ etc. This remark allows to
readily translate these Eqs. in any other reference system.

\subsection{Synchronous gauge.} \label{sec:4.2}

The synchronous frames are defined by imposing the gauge conditions:
$\phi_s = B_s = 0$. It is well known that these conditions do not fix
completely the coordinate system. Indeed, under the coordinate
transformations
\begin{equation}
\eta_s \to\tilde{\eta}_s =\eta +{\lambda\over a} ,\quad x^i \to  \tilde x^i
= x^i + \lambda^{,i} \int {d\eta\over a} \label{eq:4.5}
\end{equation}
where $\lambda$ is an arbitrary function of the spatial coordinates, the
gauge remains synchronous ($\tilde{\phi}_s = \tilde{B}_s = 0$) and (see
eqs. (\ref{eq:2.2})).
\begin{equation}
\psi_s \to \tilde{\psi}_s = \psi_s + {{\cal H} \lambda\over a}\ ;\ E_s \to
\tilde{E}_s = E_s - \lambda \int {d\eta\over a}\ . \label{eq:4.6}
\end{equation}
The function $\lambda$ and the constant of integration in (\ref{eq:4.6})
correspond to the so--called fictitious modes which do not lead to any
physical inhomogeneities. The contraint eqs. (\ref{eq:4.3}) in synchronous
frames go into (see, e.g., \cite{1}):
\begin{equation}
\psi_s = - E''_s - 2 {\cal H} E'_s ,\quad \delta \varepsilon_s = {6\over
   \kappa a^2} \left[{1\over 3} \Delta (\psi_s + {\cal H} E'_s) - {\cal H}
   \psi'_s\right]\ ,  \label{eq:4.7}
\end{equation}
with the help of eqs. (\ref{eq:4.2}) , (\ref{eq:4.7}) we can easily deduce
from (\ref{eq:3.4}) the independent matching conditions in synchronous
coordinate systems on the hypersurface $\Sigma : \varepsilon_0 + \delta
\varepsilon_s =$~const,
\begin{equation}
 [\psi_s +{\cal H} E'_s]_\pm = 0 ,\quad \left[{\cal H} E'_s
  - {1\over {\cal H}' - {\cal H}^2} \left({1\over 3} \Delta (\psi_s +
  {\cal H} E'_s) - {\cal H} \psi'_s\right)\right]_\pm =0\ . \label{eq:4.8}
\end{equation}
First we note that the particular combinations of the metric variables
entering (\ref{eq:4.8}) do not depend on the fictitious modes and we can
work directly with (\ref{eq:4.8}) to find the amplitudes of the physical
modes without specifying a particular synchronous frame. Second, if we
substitute the expressions for $\psi_\ell$, $E_\ell$ in terms of
$\phi_\ell$ in (\ref{eq:4.8}) we immediatly arrive to (\ref{eq:4.4}).
So the matching conditions (\ref{eq:4.8}) are completely equivalent to
(\ref{eq:4.4}) as it should be.

We see from (\ref{eq:4.8}) that, contrarily to what is claimed in \cite{2}
(begining of \ref{sec:4}, p. 7164), the matching conditions in a general
synchronous frame can not be reduced in general to the continuity of
metric and its first time derivatives.  The reason beeing that the
hypersurface $\Sigma$ that we choose on physical grounds as the surface
$\varepsilon_0 + \delta \varepsilon_s =$~const is {\it not} necessarily
the surface $\eta =$ const.  However, using the remaining coordinate
freedom (\ref{eq:4.5}) we can pick up the particular synchronous frame
in which the hypersurface $\Sigma :  \varepsilon_0 + \delta
\varepsilon_s =$~const.  coincide with the hypersurface $\eta_s={\rm
const}.$ It is easy to see from (\ref{eq:4.7}), (\ref{eq:4.8}) that this
is achieved if we take $\delta\varepsilon_s |_\Sigma =0$, or in explicit
form
 \begin{equation}  \left.
 \left( {1\over 3} \Delta (\psi_s + {\cal H} E'_s) - {\cal H} \psi'_s
   \right) \right|_\Sigma  = 0\ . \label{eq:4.9}
 \end{equation}
Then in the particular synchronous frame defined by (\ref{eq:4.9}) the
matching conditions (\ref{eq:4.8}) reduce to the continuity of $\psi_s,
\psi'_s, E_s'$. The quantity $E_s$ itself can always be made continuous on
$\Sigma$ since there is an extra constant of integration in synchronous
frames.  These continuity conditions together with (\ref{eq:4.9})
completely fix all the constants of the problem. For the long wavelength
perturbations when the term $\Delta (\psi_s + {\cal H}E'_s)$ can be ignored,
eq.~(\ref{eq:4.9}) simplifies to $\psi'_s |_\Sigma = 0$. Since the
perturbations of the 3-velocity $\delta u^i_s$ are proportional
to $\psi'_s$ (it follows from $0-i$ Einstein eqs), we conclude that for
long wavelength perturbations the coordinate system which satisfies
(\ref{eq:4.9}) is the particular synchronous frame which is comoving
on the hypersurface $\Sigma$, that is $\delta u^i_s|_\Sigma =0.$
Finally, we would like to stress that the matching conditions discussed here
for longwave perturbations automatically imply the conservation of $\zeta$
during the transition.

\section{Examples} \label{sec:5}

\subsection{From a pure de Sitter to a dust dominated era} \label{sec:5.1}

To start with we would like to consider some artificial example to
demonstrate that we can generate physical inhomogeneities as a result a of
transition even if we start from a pure de Sitter Universe where it is
known that the scalar metric perturbation are exactly equal to zero. As we
will show the result is determined by the physics of the transition. First
we recall some useful formulae which relate the metric perturbations in a
synhronous frame to the gauge invariant gravitational potential which can be
easily obtained by integrating (\ref{eq:2.3}) if we use there the conditions
$B= \phi =0$ and take into account the fact that $\Psi =\Phi$:
\begin{equation}
 \psi_s = \Phi + {{\cal H}\over a} \int \Phi ad\eta ,\quad E_s = - \int
  {d\eta\over a} \int \Phi ad\eta\ . \label{eq:5.1}
\end{equation}
In a pure de Sitter Universe $\Phi = 0$. The constant of integration
$\lambda$ in (\ref{eq:5.1}) (for instance, $\psi_s = \lambda {{\cal H}\over
a}$) corresponds to fictitious perturbations. If we fix it to be $\lambda_0
\not= 0$ then we pick up (up to trivial time independent transformations)
a {\it particular} synchronous coordinate system from the wide class of
synchronous flames which are related one to each other by the
transformations (\ref{eq:4.5}).  Let us assume (without any physical
justification) that the transition from the de Sitter to the dust era
happens along the hypersurface $\Sigma : \eta_0 = {\rm const.}$ in this
particular reference frame characterized by $\lambda_0$. The solution
for the gravitational potential in a dust dominated universe ($a_m
\propto\eta^2_m$) is (see(\ref{eq:2.5}))
\begin{equation}
\Phi = D_m + {S_m\over \eta_m^5} \ , \label{eq:5.2}
\end{equation}
where both constants of integration $D_m$ and $S_m$ correspond to
physical (nonfictitious) modes. Then using (\ref{eq:5.1}) one can easily
derive the solutions for the metric perturbations in the synchronous frame
during the de Sitter stage:
\begin{equation}
\psi_s = \lambda_0 H_{d.s} ,\quad E'_s = -{\lambda_0\over a}\ ,\label{eq:5.3}
\end{equation}
where the Hubble constant $H_{d.s} = {{\cal H}\over a}$ does not depend on
time in a pure de Sitter Universe, and correspondingly in the dust
dominated era:
\begin{equation}
\psi_s = {5\over 3} D_m + {2F_m\over \eta^3_m}\ ,\ E'_s = -{
D_m\over 3} \eta_m + {1\over 2} {S_m\over \eta^4_m} - {F_m\over \eta^2_m}
    \ ,  \label{eq:5.4}
\end{equation}
where $F_m$ is the constant of integration corresponding to the fictitious
mode. From the continuity conditions for $\psi_s$, $\psi'_s$ and $E'_s$
which imply the continuity of the metric and extrinsic curvature if $\Sigma
: \eta=$~const. (the continuity of $E_s$ fixes only the extra constant of
integration in (\ref{eq:5.2})) one immediatly gets that
\begin{equation}
  D_m = {3\over 5} \lambda_0 H_{d.s}\ ,\quad
 \left( {S_m\over \eta^5_m} \right)_{\Sigma +
0} = -{3\over 5} \lambda_0 H_{d.s}\ ,\quad F_m = 0\ . \label{eq:5.5}
\end{equation}

Since ${\cal D}_m$ corresponds to physical inhomogeneities we see that as a
result of the transition from a pure de Sitter to a matter dominated
Universe one produces real scalar perturbations. Of course immediatly
after de Sitter stage the gravitational potential is still zero:
$\Phi_{\Sigma + 0} = 0$, but it consists of two modes, one which decays,
while the other survives. The final result for the amplitude of the
physical perturbations is determined by the hypersurface of decay of the de
Sitter Universe (by $\lambda_0$ in example considered here).

Now, in a pure de Sitter space there is nothing which could help us to fix
the hypersurface of transition to a matter dominated era. For instance,
any hypersurface is an hypersurface of constant energy. So the physical
result is uncertain. Therefore the small deviations from a pure de Sitter
stage play a crucial role in predicting the resulting inhomogeneities,
since they actually define the physics of the transition.  The most
natural possibility (in the absence of further concrete model) is to
take as $\Sigma$ the hypersurface of constant energy density as we did.

\subsection{From inflation to a radiation and then dust dominated era}
\label{sec:5.2}

 Now we consider the model studied in \cite{2}. Namely, we assume that
the Universe went through three stages: at first there was inflation
$(i)$ with scale factor $a_i \propto (-\eta_i)^{-1-\gamma}$, where
$0 \ll\gamma\ll 1$, then a radiation dominated era $(r)$ with $a_r \propto
\eta_r$ and finally a dust matter dominated stage $(m)$: $a_m \propto
\eta_m$. Let us denote the times of transition from $i$ to $r$ stage as
$\eta_1$ and correspondingly from $r$ to $m$ stage as $\eta_2$
in terms of conformal time $\eta_r$ in the $r$-stage. Under the above
parametrization of the scale factors, conformal time $\eta$ jumps during
the transitions and to relate, for instance $\eta_i$ to $\eta_r$ at this
point we can use the continuity condition for ${\cal H}$. The time $\eta$
can be easily made to be continuous by shifting $\eta_r$ and $\eta_m$ by
constant factors. However it has no impact on the final results and to
simplify the formulae we prefer to use the parametrization above. Also
from the very beginning we restrict ourselves only to inhomogeneities
with scales bigger than the Hubble radius at $\eta =\eta_2$, that is
$(k\eta_2) \ll 1$, where $k$- is the comoving wavenumber of planewave
perturbations. Then for such perturbations we can write the solutions
for the gravitational potential $\Phi$ in the long wavelength
approximation during the inflation and radiation dominated stages as:
\begin{mathletters}
 \label{eq:5.6}
\begin{eqnarray}
 \Phi_i &\simeq & \sqrt\gamma H^k_i \ , \label{eq:5.6a} \\
 \Phi_r &\simeq & D_r \left ( 1 - {(\omega \eta_r)^2\over 10}
\right) + {S_r\over \eta^3_r}\ . \label{eq:5.6b}
\end{eqnarray}
\end{mathletters}
We used Planck units and
$H^k_i$  is the value of Hubble constant at the moment of horizon
crossing by the perturbation with the wavenumber $k$ during inflation.
Imposing the perturbation to be in their quantum grond state during
inflation fixes the amplitude in (\ref{eq:5.6a}) \cite{4} (there is no
disagreement in the literature on this point) and we ignored the decaying
mode which will be completely irrelevant at the end of inflation. In the
radiation stage we keep in (\ref{eq:5.6b}) the first correction $\sim
(\omega\eta)^2$, where $\omega = k/\sqrt 3$, to the constant
nondecaying mode since at leading order this mode has a vanishing
derivative.

The solution (\ref{eq:5.6b}) can be obtained from well known exact
solution \cite{1,2} in the $r$-stage, if we expand it in powers of
$\omega\eta$. Using (\ref{eq:5.6}) one easily gets with the help of
(\ref{eq:5.1}) the following expression for $\psi_s$ and $E'_s$ in a
synchronous frame:
\begin{eqnarray}
  \psi_i &=& {1+ 2\gamma\over \sqrt\gamma} H^k_i + (1+\gamma)
   F_i (-\eta_i)^\gamma\ , \nonumber \\
   E'_i &=& - {1\over \sqrt \gamma} H^k_i (-\eta_i) - F_i
   (-\eta_i)^{1+\gamma} \ , \label{eq:5.7}
\end{eqnarray}
in the inflationary stage $(i)$ and
\begin{eqnarray}
  \psi_r &=& {3\over 2} \left( 1 - {1\over 12} (\omega\eta_r)^2\right)
  D_r + {F_r\over \eta_r^2} \ , \nonumber \\
  E'_r &=& - {1\over 2} \left( 1 - {1\over 20} (\omega\eta_r)^2\right)
 \eta_r D_r +{S_r\over \eta_r^2}-{F_r\over \eta_r}\ , \label{eq:5.8}
\end{eqnarray}
in the $r$-stage, where the constants of integration $F_i$, $F_r$
correspond to fictitious modes.

The potentials in a dust matter dominated Universe were given before (see
formulae (\ref{eq:5.2}), (\ref{eq:5.4})). Now if we claim that $\psi$,
$\psi'$, $E'$ are continuous two on the transition hypersurfaces, $\eta_r =
\eta_1$ and $\eta_r = \eta_2$, then we get 6 eqs. for seven unknown
constants $D_r$, $S_r$, $D_m$, $S_m$, $F_i$, $F_r$, $F_m$.
Clearly we need one extra condition. The one imposed in \cite{2} is that
in the dust dominated era the synchronous coordinate system be also
the comoving one. This fixes the constant $F_m$ to be zero. Then a
straightforward calculation gives us the following result for the constant
$D_m$:
\begin{equation}
   D_m = {3\over 5} {(1+2 \gamma)\over \sqrt\gamma} \left( 1+
  {2+\gamma\over 12\gamma} (\omega\eta_m)^2 (\eta_2/\eta_1)^2\right)^{-1}
  \label{eq:5.9}
\end{equation}
which agrees with \cite{2}.

Recall that $D_m$ is the gravitational potential in the $m$-stage. Since
$\gamma \ll 1$ we see from (\ref{eq:5.9}) that for cosmologically relevant
scales for which ${1\over\gamma} (\omega\eta_2)^2 (\eta_2/\eta_1)^2
\gg 1$, the spectrum (\ref{eq:5.9}) is by no means scale invariant (for
which $D_m$ doesn't depend on $k$).  We do not agree with the author of
paper \cite{2} who concluded on the base of consideration similar to the
above one that it implies a scale invariant spectrum in relevant for
COBE scales.
The reason for disagreement is that he took, as we think mistakenly, in
formula for ${\cal D}$ on the page 7166 \cite{2} the parameter $\gamma_1$
to be equal 2 (it corresponds to $\gamma =-2$ in our notations) instead
of $\gamma_1\ll 1.$ Since the spectrum (\ref{eq:5.9}) doesn't resemble at
all the standard result (see \cite{1}) it is quite interesting to find
out the reason for this desagreement.

As we argued in Sect.~\ref{sec:3}, the physics of the transition tells
us that it should happen on the hypersurface of constant energy. This
implies that, for instance, in an arbitrary synchronous coordinate
system the matching conditions should be (\ref{eq:4.8}). They are
equivalent to the continuity conditions for $\psi_s$, $\psi'_s$, $E'_s$
only in the particular synchronous frame which for long wavelength
perturbations is close to the comoving one at the moment of transition.

The precise condition which fixed this coordinate system and,
accordingly, the amplitudes of the fictitious modes $F$ is given by
(\ref{eq:4.9}). If we decide the synchronous frame to be comoving one in
the $m$-stage $(F_m=0)$ then this coordinate system satisfies the
requirement (\ref{eq:4.9}) with very good accuracy for $k\eta_2\ll
1$ on the hypersurface of transition from radiation to matter dominated
era. The function $F_r$ which specifies this particular synchronous frame
II in the $r$-stage is $^{\rm II}F_r \sim \omega^2 (\eta_2)^4 D_r$.
However, the frame II is very far from the comoving one on the hypersurface
of the first transition from the $i$ to the $r$ stage. Actually in the
frame I which is comoving on the hypersurface of the first transition the
amplitude of the fictitious mode is $^{\rm I}F \sim \omega^2 (\eta_1)^4
D_r$.  Since $^{\rm II}F/^{I}F \sim \eta_2/\eta_1 \gg 1$, we see that
the frames I and II are very different. As one can check this difference
is quite important for the perturbations with $(\omega \eta_2)^2
(\eta_2/\eta_1)^2 \gg 1$. How then should we proceed to get the correct
answer for the final amplitude of scalar perturbations~? The simplest
way is just to use the matching conditions (\ref{eq:4.8}) which do not
involve the fictitious modes $\propto F$ and permit us immediately to write
4 eqs. for 4 unkonwn constants $D_r$, $S_r$, $D_m$, $S_m$.
 However, if we want to insist on the continuity conditions for
$\psi_s$, $\psi'_s$ and $E'_s$ then we can do it {\it but}, after the
matching in the particular frame I at $\eta_r =\eta_1$ we should go
to the {\it other} synchronous frame II to make the matching at $\eta_r =
\eta_2$. This procedure fixes all the constants of the problem. As
one can easily check both ways lead to the standard result for the final
amplitude of scalar perturbations:
 \begin{equation}
  D_m = {3\over 5} {H^k_i\over \sqrt\gamma}\ , \label{eq:5.10}
 \end{equation}
for $\gamma\ll 1$, which is very different from (\ref{eq:5.9}), while
for longwave gravity waves $h\simeq H^k_i$.

Finally there is noting strange about the fact that the final amplitude of
scalar perturbations can go to infinity when $\gamma \to 0$ despite the
fact that perturbations during inflation go to zero (see (\ref{eq:5.6a})).
As it was shown in Sect.~\ref{sec:5.1} as a result of the decay of a pure
de Sitter stage we can get whatever we want unless the physics of
transition is specified. In the model considered above the hypersurface
of transition becomes more ``inhomogeneous" (we mean its extrinsic
curvature) when we approach a pure de Sitter space. The formulae
(\ref{eq:5.10}) has however a limited range of validity since it was
obtained in the {\it linear} perturbation theory. Thus we can believe it
only if $\gamma \gg H^2_I$ $(\Phi \sim D \ll 1)$.
We hope that the above consideration clearly demonstrates that the
calculations which are very straightforward in terms of gauge invariant
variables (or in the longitudinal gauge) can become sophisticated
in synchronous frames.

\subsection{Fate of the ``decaying'' mode (applications to string cosmology)}
\label{sec:5.3}

 Finally we want to consider some consequences of the obtained results
for string cosmology. It is well established (see for instance, \cite{13})
that in the models with a dilatonic field in the absence of
nonperturbative potential there is a stage in the evolution of the
Universe which is described by ``superinflationary'' expansion in the
``Brans-Dicke frame''. It is usually assumed that this stage should
serve the same purposes as ordinary inflation \cite{13}. In the ``Einstein
frame'' this superinflationary stage corresponds to a contracting
Universe and the ``decaying'' mode in (\ref{eq:2.5}),
 \begin{equation}
  \Phi = S {H\over a} \ , \label{eq:5.11}
\end{equation}
(we consider only long wavelength inhomogeneities) grows in this
contracting Universe \cite{14}. There is therefore the danger that this mode
will eventually turn in to a nondecaying mode and create big
inhomogeneities in the final Friedmann Universe thus completly
invalidating the original model. It was found in \cite{14} that the mode
(\ref{eq:5.11}) has peculiar properties which give us the hope to
avoid this danger. However, the question of what will happen with it
after the transition from the superinflationary to the Friedmann stage
was not clarified. The problem is that there are no realistic models
which solve the ``graceful exit'' problem and allow us to go smoothly
from the superinflationary to Friedmann era \cite{15}.
In the absence of such a model we will assume that the transition happens
as a result of a change in the effective equation of state along a
hypersurface of constant energy. On the basis of the results obtained
above it turns out to be enough to predict the fate of the mode
(\ref{eq:5.11}) (which is growing in superinflationary Universe) after
transition to Friedmann era.

As we have seen before the nondecaying mode of the gravitational potential
in an expanding Universe after the change in the equation of state is
redistributed in comparable proportions between nondecaying and decaying
modes. If it would be also the case for the decaying mode the situation
would be quite unfavorable for the superinflationary models. However
substituting (\ref{eq:5.11}) in (\ref{eq:4.4}) we immediately find that
 \begin{equation}
   S^+ = S^- \label{eq:5.12}
\end{equation}
to leading order in $(k\eta)$-expansion. A more careful analysis shows
that the nondecaying mode will be generated only at the next order in
$(k\eta)$, that is $D^+\simeq (k\eta)^2 S^-$. Thus we see that the
growing mode of scalar perturbations in superinflationary Universe will be
practically entirely converted in the decaying mode in Friedmann Universe.
This result is general and doesn't depend on the concrete model for
transition from superinflationary to Friedmann era. The only restriction
is that the transition happens on a hypersurface of constant energy.

\section{Discussion} \label{sec:6}

 We derived the matching conditions for cosmological perturbations
during a sharp change in the equation of state in gauge invariant form
and in different gauges (including the synchronous one) assuming that the
transition from one stage to the other happens on the hypersurface of
constant energy. This last restriction is quite natural and it is
dictated by the physics of the transition. Using it we confirmed and
clarified the standard result concerning the ratio of density
perturbation to gravity waves predicted by inflationary models.

 We showed that the matching conditions used in \cite{2} do not satisfy
this requirement and therefore lead to results for the spectrum of
density perturbations at odds with the standard results.

 The results obtained go beyond the model usually considered and can be
applied also study to the fate of cosmological perturbations in models
where not all of the details of the physics of the transition are specified.
The only important piece of physics which we used is that the transition
happens on a hypersurface of constant energy. The results were applied
to follow the fate of the growing mode in superinflationary cosmology .

A final remark which we would like to make concerns the ``best'' gauge
for cosmological perturbations. As far as we are aware the question
about ``what is the best'' gauge or the advantage of the gauge invariant
formalism is still under debate in the literature. Our (subjective)
point of view is quite simple. There are of course no arguments of
principle which would permit us to decide: all gauges, and even the
gauge invariant formalism are on the same footing; one can work in any
gauge.  We hope however that in this paper we clearly demonstrated that
for practical purposes the gauge invariant formalism (or equivalently
the longitudinal gauge) is by far the most convenient one for treating
cosmological perturbations and is an insurance against mistakes.

\acknowledgments

We thank L.P.  Grishchuk for raising our interest in this problem and
comments and C.~Gundlach, D.~Langlois, J.~Martin, D.~Polarski and
C.~Schmid for discussions.  V.F.~Mukhanov would like to thank the
D\'epartement d'Astrophysique Relativiste et de Cosmologie for
hospitality and the CNRS for financial support.

\end{document}